# Gate-tuneable single-photon emitters in WSe$_2$ monolayer created via AFM nanoindentation on rigid SiO$_2$/Si substrates

*[1†]Ajit Kumar Dash, [2†]Sanket Jugade, [2]Manavendra Pratap Singh, [1]Hardeep, [3]Tilly Guyot, [3]Cora Crunteanu-Stanescu,[4]Indrajeet Dhananjay Prasad, [4]Yunus Waheed, [4]Sumitra Shit, [3]Sébastien Roux, [4]Santosh Kumar, [3]Cedric Robert, [3,5]Xavier Marie, [2*]Akshay Naik, [1*]Akshay Singh*

[1]Department of Physics, Indian Institute of Science, Bengaluru, Karnataka 560012, India

[2]Centre for Nanoscience and Engineering, Indian Institute of Science, Bengaluru, Karnataka 560012, India

[3]Université de Toulouse, INSA-CNRS-UPS, LPCNO, 135 Avenue de Rangueil, 31077 Toulouse, France

[4]School of Physical Sciences, Indian Institute of Technology Goa, Ponda, 403401, Goa, India

[5]Institut Universitaire de France, 75231 Paris, France

† Authors contributed equally to the work

*Corresponding authors: anaik@iisc.ac.in, aksy@iisc.ac.in

**Abstract:** Single-photon emitters (SPEs) hosted by two-dimensional (2D) semiconducting materials are envisioned for next-generation quantum applications. However, SPE creation in 2D semiconductors on rigid substrates like SiO$_2$/Si via nanoindentation is a technological gap, critical for interfacing SPEs with photonic circuits and cavities. Here, we report a protocol for deterministically creating SPEs in monolayer WSe$_2$ on SiO$_2$/Si substrates using a sharp diamond AFM (atomic force microscope) tip. A displacement-controlled indentation process is developed, allowing indent depths > 150 nm necessary for creating SPEs. Sharp defect peaks (~200 μeV) are observed in cryogenic (4K) photoluminescence (PL) spectrum at nanoindented sites and are stable upto ~ 120K. 76% of sites exhibit sharp defect-bound peaks confirmed by power-dependent, temperature-dependent, and time-resolved PL (TRPL). AFM and PL mapping link these peaks to indent periphery. The peaks show sub-linewidth spectral jitter, no blinking, and single-photon nature in second-order autocorrelation measurements. SPEs can be switched on/off, and background emissions suppressed using electrical gating. Gate-voltage dependent TRPL indicate that SPE dynamics can be tuned, depending on nature of SPE, pointing the way to higher-purity SPEs. Our work is directly applicable to other 2D materials and photonic circuit/cavity compatible rigid substrates and is a significant step for scalable SPE technologies.

**Introduction:**

Solid-state single-photon emitters (SPEs) are essential components of modern quantum technologies, including quantum computing, communication, and sensing[1,2]. SPEs are characterized by several parameters including single-photon purity, indistinguishability, brightness, reliability, gate-tunability, and excitation efficiency[1,3]. Further, spin-photon coupling, and beyond-cryogenic (4 K) temperature operation are required for SPE applications. Moreover, creation of reliable, spatially deterministic SPEs that can be integrated with practical quantum devices are critical for widespread adoption of SPEs[3,4]. SPEs are commonly realized in quantum dots, color centers in bulk materials including diamond and silicon carbide (SiC), and localized defects in two-dimensional (2D) materials[1].

The atomic thickness and planar structure of 2D materials makes them ideal for compact devices and enable easy integration with quantum photonic platforms[5]. SPEs hosted by 2D materials offer versatility and tunability in the photon emission energies, and have excellent SPE characteristics[3,5–7]. Specifically, SPEs in transition metal dichalcogenides (TMDs) exhibit narrow spectral linewidth and high single-photon purity under cryogenic conditions[1,3]. Defect and strain engineering of monolayer (ML) TMDs can create bound exciton manifolds, which are ideal for SPEs. However, SPEs in TMDs are usually randomly positioned[8,9]. To enhance brightness, it is necessary to integrate SPEs within low mode volume cavities that possess large Purcell factors.

Among TMD members, ML $WSe_2$ is reported to host the highest purity SPEs[10–12]. Strain engineering using nanopillars has been explored for creation of SPEs in $WSe_2$[13]. However, coupling of nanopillar or nanosphere based SPEs to photonic circuits and cavities is complex. Achieving deterministic creation of SPEs in 2D materials after transfer onto desired photonic structures thus becomes essential.

Atomic force microscopy (AFM) nanoindentation is a promising solution for deterministic creation of SPEs after transfer of 2D materials onto substrates. So far, nanoindentation has been demonstrated only on soft polymer surfaces such as polymethyl methacrylate (PMMA) that constraint the design of photonic chips/cavities[10,14]. Further, polymers can degrade SPE properties over time[15]. Compared to polymer substrates, $SiO_2$/Si substrates are thermally and mechanically stable, and are compatible for photonic chips and cavities. Thus, nanoindentation of ML $WSe_2$ directly on $SiO_2$/Si substrates will pave the way for scalable integration of SPEs into practical quantum devices.

This work uses a sharp diamond AFM tip to perform nanoindentation in ML $WSe_2$ directly on $SiO_2$/Si substrate. A displacement-controlled AFM nanoindentation method is developed, enabling precise, consistent indentations with either constant or varying depths across multiple locations. Cryogenic (4 K) photoluminescence (PL) spectrum acquired from the indented sites with depths greater than 150 nm show sharp bound-exciton peaks with peak widths less than a few 100 μeV. At 4 K, most of the sharp peak's emission is contributed by the zero-phonon line (ZPL). The sharp peaks persist up to 120 K, indicating excellent thermal stability. Cryogenic PL mapping demonstrates high SPE yield and reproducibility of the nanoindentation process. Further, correlating PL map with the AFM image identifies periphery of the nanoindent as the spatial origin of sharp peaks. Power-dependent PL and time-resolved PL (TRPL) experiments indicate bound nature of the sharp exciton peaks. The sharp defect peaks show sub-linewidth spectral stability in jittering experiments, with no blinking. The single-photon nature of emission is confirmed by second-order autocorrelation ($g^{(2)}$) measurements. Finally, we fabricated gated devices on $SiO_2$/Si substrates and investigated carrier doping dependence of ML $WSe_2$ SPEs. Gate voltage can be used to turn the emitters on or off, suppress background PL emission, and tune SPE dynamics in TRPL measurements. Thus, this gate-tuneable platform allows the identification and operation of high-purity SPEs in 2D materials.

**Results and discussion:**

**AFM Nanoindentation of ML $WSe_2$ on $SiO_2$/Si substrate**

ML $WSe_2$ flakes were exfoliated from bulk crystals onto $SiO_2$/Si substrates using a PDMS-assisted blue tape technique (see Methods). Nanoindentation on freshly exfoliated ML flakes was carried out using AFM. Prior nanoindentation studies with $WSe_2$ supported on soft polymer substrates typically use an AFM probe with a silicon tip mounted on a compliant cantilever[14]. This earlier approach, however, is unsuitable for rigid substrates like $SiO_2$/Si, due to their substantially higher elastic modulus and hardness (c.f. polymer substrates). To address this challenge to indent monolayer WSe$_2$ on a SiO$_2$/Si substrate, we used a sharp diamond tip (~ 10 nm radius) mounted on a stiff cantilever (~ 350 N/m) (see Methods).

Initial nanoindentation experiments were conducted in force-controlled mode, allowing maximum indent depths of up to ~ 30 nm under an applied load of ~ 20 µN (Supporting Information section - I). The limitation on maximum achievable depth arises from the AFM system's optical detection, where increased cantilever curvature at higher loads causes the laser reflection to move out of the four-quadrant photodiode. This force-controlled approach led to PL modulation at room temperature at the indentation sites; however, no sharp PL emission was observed at cryogenic temperatures (refer to Figure S1d).

To overcome the depth limitation, we adopted displacement-controlled indentation, wherein the Z-piezo displacement is precisely controlled to drive the AFM tip into the substrate to a target depth (Figure 1a). To accurately achieve the targeted indentation depth, calibration of the Z-piezo displacement was performed using indentation on bare SiO$_2$/Si substrate (see Methods, and Supporting Information section - II). Then, systematic depth-dependent indentations on a single WSe$_2$ flake revealed that indent depths > 150 nm are required to create sharp defect peaks (Supporting Information section - III). To demonstrate the depth reproducibility of this process, an array of 41 indentations was created on another ML WSe$_2$ flake with a target depth between 210 and 220 nm, as shown in the AFM topography micrograph in Figure 1b. Although this large-area topography micrograph helps visualize the indented sites across the flake, the individual indent depth was determined by small-area high-resolution AFM imaging at the corresponding site (see Methods). The average indent depth from measurements across all indented sites on this flake was 213.6 nm with a standard deviation of 3.4 nm, highlighting the excellent precision and repeatability of this technique. In this way, throughout the study, we achieved either uniform target-depth or controlled, depth-varying indentations on a single flake using the diamond AFM tip.

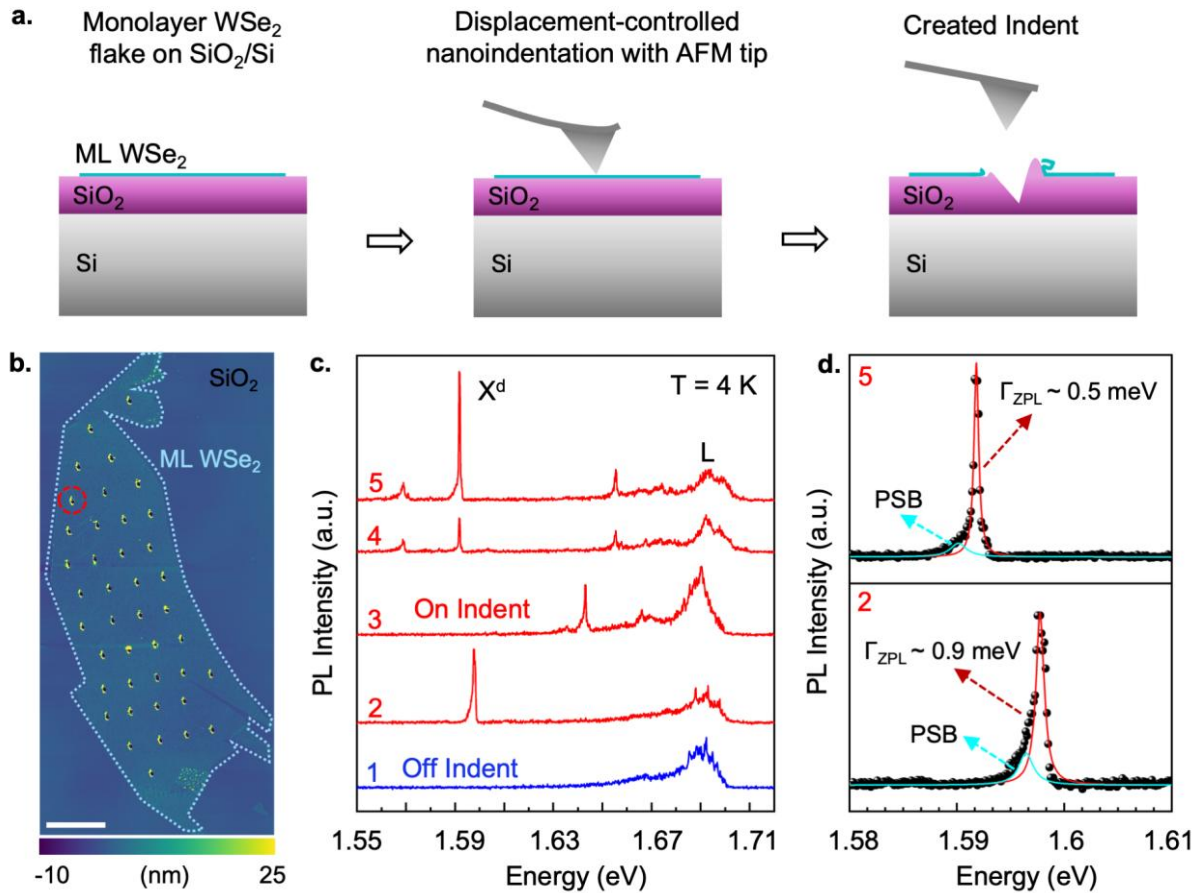

**Figure 1: AFM nanoindentation of monolayer (ML) WSe$_2$ on SiO$_2$/Si substrate. (a)** Schematic of the displacement-controlled AFM nanoindentation process of ML WSe$_2$ flake on SiO$_2$/Si substrate. **(b)** AFM topography micrograph of a large-area ML WSe$_2$ flake with equally spaced 41 indents created with constant target depth. Here, the height scale range has been narrowed for better visibility of indent sites; however, the average depth of indents is ~ 213 nm. Horizontal scale bar in Figure (b) is 10 μm. **(c)** Representative cryogenic (4 K) PL spectra acquired from a few nanoindented sites and pristine region of the flake. **(d)** Lorentzian peak fitting of two sharp peaks (X$^d$) showing zero phonon line (ZPL), and phonon side band (PSB) on the lower energy side.

**Cryogenic (4 K) PL characterization of nanoindented ML WSe$_2$**

Cryogenic (4 K) PL spectroscopy enables precise characterization of SPEs created via AFM nanoindentation. Representative cryogenic (4 K) PL spectrum acquired from few nanoindented sites are shown in Figure 1c. Spectrally sharp PL peaks (X$^d$) are observed in the energy range 1.5 – 1.7 eV, where SPEs in ML WSe$_2$ usually appear[8,14]. The sharp peaks were absent in the PL spectrum obtained from off-indent sites (bottom spectrum of Figure 1c), which suggests that sharp peaks were created due to nanoindentation process. Additionally, a broad PL peak (L peak) appears around 1.7 eV, which is usually referred to as classical background, and is present in both indented and non-indented regions[10,16]. The L peak is also observed in non-indented samples, and is associated with the recombination of localized exciton peaks[17–19]. The effect of L peak on quantum properties of lower energy sharp peaks is discussed in the next sections.

The sharp peaks are asymmetric in nature, consisting of a zero-phonon line (ZPL), and a phonon side band (PSB) at the lower energy side. In order to measure linewidth, Lorentzian peak fitting was performed for sharp peaks in spectra 2 and 5 of Figure 1c. The extracted linewidth ($\Gamma$) of the ZPL is approximately 0.5 meV and 0.9 meV for spectrum 5 and 2 (Figure 1d), respectively. In spectra from other samples, we have observed ZPL as narrow as 0.2 meV. The observed linewidths are comparable to reported linewidth values of SPEs in ML WSe$_2$[8,13,20]. The ZPL accounts for majority of a particular peak's emission intensity, suggesting high-quality SPEs with low inhomogeneous broadening.

**Spatial distribution and origin of sharp peaks (X$^d$)**

Cryogenic (4 K) PL mapping is essential to study the spatial distribution of the sharp defect peaks and yield of our AFM nanoindentation process on SiO$_2$/Si substrate. Figure 2 (a - c) corresponds to cryogenic PL (4 K) maps of the nanoindented ML WSe$_2$ in specific detection energy ranges (see Methods). The PL map in the energy range 1.63 - 1.68 eV (Figure 2a) shows bright spots around some of the indented sites corresponding to $X^d$ peaks, with minor PL contributions from the L peak (reference PL spectra are presented in Supporting Information section - IV). In the energy ranges 1.59 - 1.63 eV and 1.55 - 1.59 eV (Figure 2b and 2c), we observed bright spots mostly around the indented sites. These are attributed to the presence of spectrally-isolated sharp peaks and reduced intensity of L peak. Spectrally-isolated sharp peaks with reduced background are more likely to show higher SPE purity, compared to sharp peaks spectrally close to or superimposed with L peak. Further, considering all PL maps (Figure 2a - 2c), nearly 76 % of nanoindentation sites (20 out of 26) show localized PL emission, demonstrating the high yield of the AFM nanoindentation of ML WSe$_2$ on SiO$_2$/Si substrate for SPE creation.

We now examine the spatial origin of the SPEs created by the nanoindentation process by correlating the individual indent morphology and its cryogenic PL map. Figures 2d and 2e show the topography and phase shift micrographs for an indent highlighted by the yellow circle in Figure 2a. Figure 2f shows the high-resolution PL map of the same indent, obtained by integrating PL in the energy range 1.55 – 1.72 eV. By comparing the indent topography with the PL map, it becomes evident that the PL is significantly quenched at the center of the indent, indicating removed WSe$_2$ material. However, bright spots appear along the indent perimeter, specifically near the lower side of the periphery. Representative PL spectra from the center and from the lower periphery of the indent are displayed in Figure 2g. The sharp peaks are prominent at the periphery of the indent. Coincidentally, sharp peaks emerging from the indent periphery were also reported in an previous study on AFM indentation of ML WSe$_2$ on polymer substrates[21]. However, for nanoindentation of WSe$_2$ layer on polymer substrates, it is often hypothesized that the WSe$_2$ layer deforms along the indent contour[14]. Thus, to determine the spatial origin of the sharp emission in our work, it is essential to understand how AFM nanoindentation affects a 2D material on a rigid SiO$_2$/Si substrate.

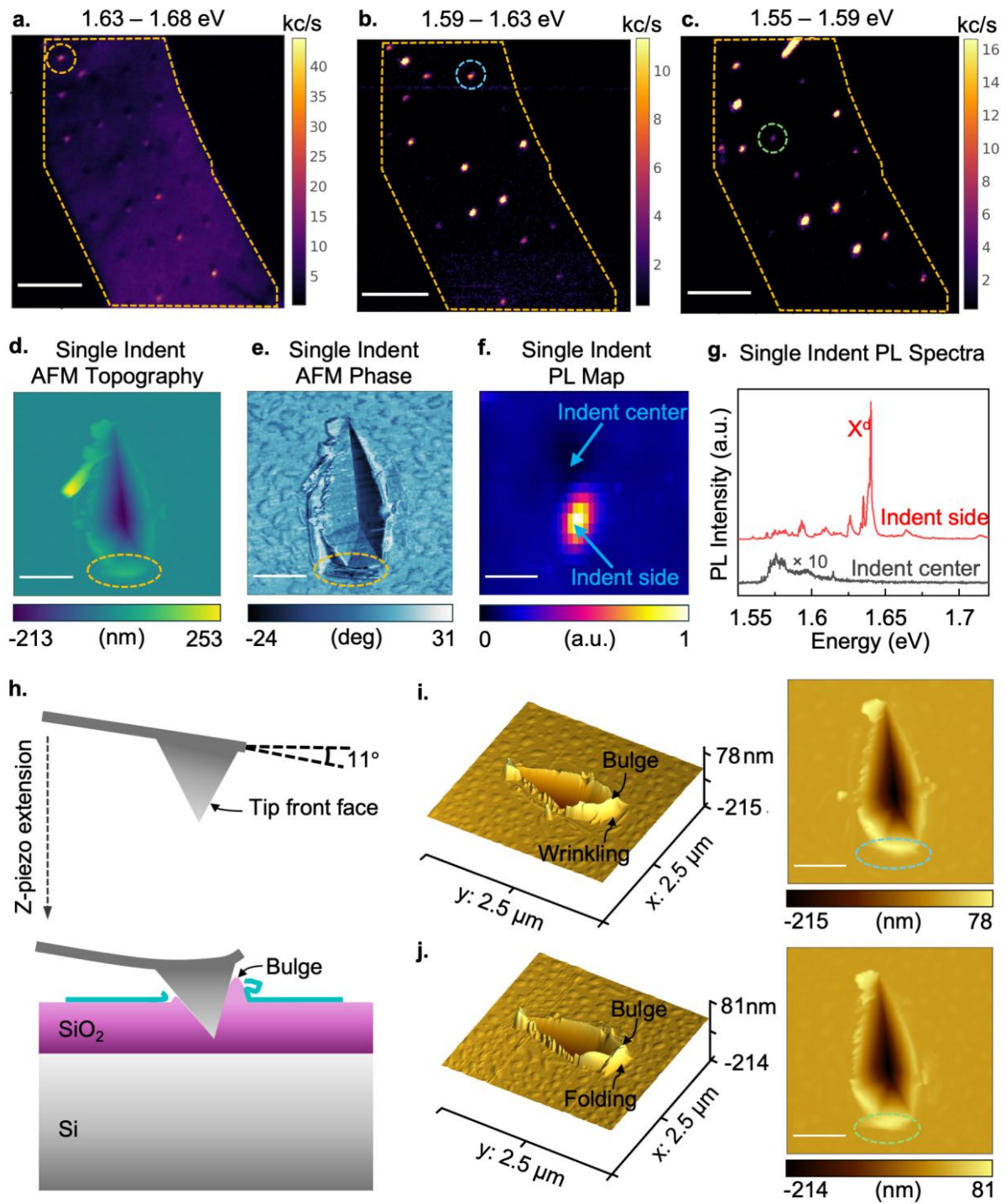

**Figure 2: Spatial distribution and origin of sharp peaks: (a - c)** Cryogenic (4 K) PL mapping of nanoindented ML WSe$_2$ in specific detection energy ranges (indicated on top of the figure). Horizontal

scale bars in (a-c) correspond to 10 μm. **(d, e)** AFM topography and phase shift micrographs of a single indent, highlighted by the yellow circle in (a). **(f)** Cryogenic PL map of the single indent shown in AFM images (d, e). Horizontal scale bars in (d-f) correspond to 500 nm. **(g)** Representative PL spectra from center and periphery of the indent PL map shown in (f). **(h)** Schematic showing how the ~ 11° cantilever mounting angle and pronounced cantilever curvature during deep $SiO_2$ indentation produce a substrate bulge at the front face of the AFM tip and lead to localized out-of-plane deformation of the torn flake edge at the bulged region. **(i, j)** 3D projection of the AFM topography together with topography micrographs of two different indents highlighted by the blue circle in (b) and the green circle in (c), showing localized wrinkling and folding of the torn edge of $WSe_2$ at the bulged region. Horizontal scale bars in (i, j) correspond to 500 nm.

The AFM topography micrograph (Figure 2d) reveals an indent depth of 213 nm, along with features around the indent periphery that resemble flake tearing. The corresponding phase-shift micrograph (Figure 2e) shows distinct phase contrast along the right and left edges of the indent, confirming the presence of torn flake edges. The phase contrast also confirms that the $WSe_2$ layer is absent within the indent. In other words, the $WSe_2$ layer does not deform along the indent contour. Instead, it is sculpted within the indented region likely due to shear-induced tearing during indentation. Similar nanoindentation-induced sculpting of monolayer $WS_2$ and $MoS_2$, and graphene on $SiO_2$/Si substrates has been reported using instrumented indenters operating at millinewton-scale loads, with indent depths ranging from a few nanometers up to 100 nm[22–24].

Beyond sculpting of the flake, the indent morphology also shows a distinct bulge at the lower side of the indent periphery, with a peak height of ~ 78 nm (Figure 2d). This bulge arises from large and preferential pile-up of displaced $SiO_2$ material during indentation. Comparing the AFM images and PL map reveals that the sharp peaks originate from near this bulged region.

The circled areas of indent in Figure 2d and 2e show localized folding of the torn flake edge near the bulged region, which may play a critical role in creating the sharp defect peaks.

To further probe the origin of the sharp peaks, we performed additional indentations of WSe$_2$ on SiO$_2$/Si using an instrumented nanoindenter, with indent depths ranging from 79 nm to 271 nm (see Supporting Information section - V). Interestingly, no sharp peaks were observed from these indented sites in the cryogenic (4 K) PL measurements. AFM images of indents created using the instrumented indenter and AFM nanoindentation showed sculpting of flake in both cases, with no WSe$_2$ material present inside the indented region. However, the indents created using the instrumented indenter exhibited a symmetric shape and lacked the prominent bulge typically observed at the lower side of the indent periphery of AFM-created indents. This suggests that the characteristic substrate bulge observed only in AFM indentation may be a key factor in creating the sharp defect peaks.

The pronounced bulging of the SiO$_2$/Si substrate observed in AFM nanoindentation, which is absent in instrumented nanoindentation, arises from fundamental differences in the implementation of these two indentation methods. In an instrumented indenter, the tip is mounted on a stiff vertical shaft that approaches the sample surface perpendicularly, resulting in a uniform material pile-up around the indent. In contrast, the AFM cantilever is mounted at a small tilt angle so that the tip contacts the sample before the cantilever body and ensures proper laser reflection onto the detector. In our setup, the cantilever was mounted ~ 11° relative to the horizontal plane. This geometry, combined with large vertical deflection of the cantilever (beyond 10 V saturation of photodiode) at large indentation depths, leads to significant, preferential pile-up of SiO$_2$ along indent boundary on the front face of the tip (Figure 2h). Additionally, during the initial phase of indentation at smaller indent depths itself, shear forces tear the flake. As the tip then continues to plastically deform the SiO$_2$ substrate, the resulting bulge compresses the torn flake edge and induces restructuring in the form of localized folds/creases (as observed in Figure 2d and 2e) or wrinkles in the flake near the bulged region. Figure 2i and 2j show 3D projections of the topography for two different indents (circled in Figures 2b and 2c, respectively), which help visualize the localized wrinkling and folding of the flake edge induced by substrate bulging during the indentation process.

Therefore, the characteristic bulging of the $SiO_2$ substrate at the front face of the AFM tip, together with the local out-of-plane deformations of the torn $WSe_2$ flake edge near the bulge, produces a unique combination of defect states and localized strain. This mechanism is central to the origin of the SPEs in our method.

**Bound nature and spectral stability of the sharp peaks ($X^b$)**

In order to classify the sharp peaks as bound excitonic states, we performed excitation laser power-dependent PL. The integrated intensity of the sharp PL peak ($X^d$) shows saturation behaviour with laser power (Figure 3a), a signature of defect-bound excitons with a low density of defect states. For a two-level system, the saturation powers ($P_{sat}$) for finite density of single defects can be obtained by fitting the data with the equation

$$I = k \frac{P}{P + P_{sat}}$$

I is the PL intensity, P is the laser power, and k is a fitting constant. The calculated saturation power ($P_{sat}$) for the shown $X^d$ peak is ~ 44 μW (laser spot size ~ 1.3 μm). These values are comparable to the saturation powers for SPEs observed in $WSe_2$[8,14]. Further, the L peak has sublinear laser power dependence, indicating its localized nature[25,26]. The detailed power coefficient analysis of sharp, L, and neutral exciton peaks is presented in Supporting Information section - VI.

TRPL measurement was employed to measure lifetimes of the sharp peaks. TRPL was performed using a 710 nm pulsed laser excitation with 8 MHz repetition rate (see Methods). TRPL of sharp peak ($X^d$) shows a single exponential decay, with lifetime ($\tau_{Xd}$) ~ 5 ns (Figure 3b, fitting of TRPL data is shown in Supporting Information section - VII), similar to the reported lifetime values of strain induced SPEs in ML WSe$_2$[14]. The representative PL spectrum acquired from nanoindented site is shown in the inset of Figure 3b. Interestingly, TRPL measurement of L peak indicates a biexponential decay with time constants 0.6 ns and 20 ns, indicating its different origin compared to sharp peaks (Supporting Information section - VII).

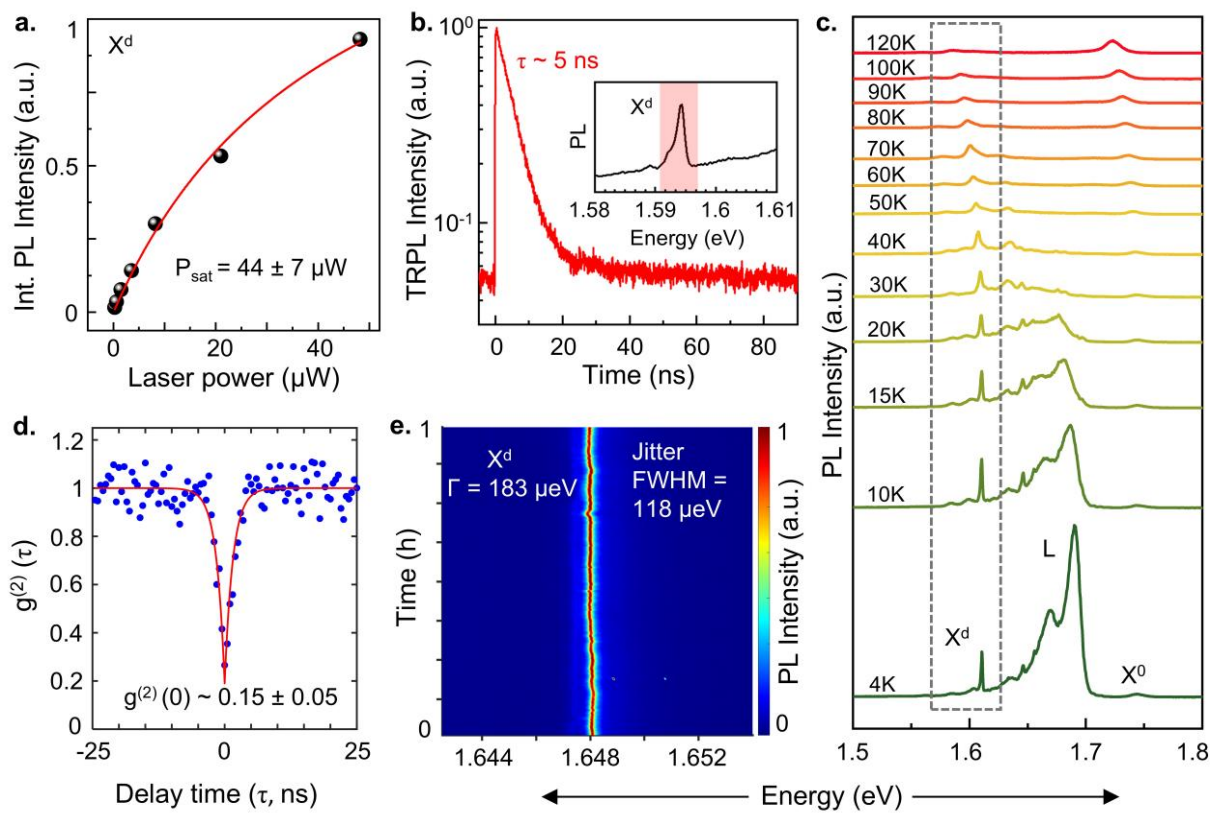

**Figure 3: Bound nature and spectral stability of sharp peaks. (a)** Integrated PL intensity of sharp peak ($X^d$) plotted against laser power and fitted to power saturation law. **(b)** Time-resolved PL (TRPL) spectrum of sharp peak ($X^d$) plotted against delay time. **(c)** Evolution of sharp peaks with temperature. **(d)** Representative second-order autocorrelation ($g^{(2)}(\tau)$) measurement for sharp peaks showing antibunching dip. **(e)** Jittering of sharp peak measured via surface plot of kinetic series consisting of 3600 PL spectra, with each spectrum acquired for 1s.

Importantly, the bound sharp peaks persist up to 120 K temperature (Figure 3c), pushing the limit towards higher temperature SPEs[27]. The intensity of these sharp peaks quenches with temperature, consistent with the bound exciton model[28]. On the other hand, we observed increasing intensity of neutral exciton ($X^0$) peak with increasing temperature, attributed to thermal depopulation of the dark ground state of ML WSe$_2$[29]. Both the $X^0$ and $X^d$ peaks redshift with increase in temperature, owing to temperature dependence of the bandgap of ML WSe$_2$[30]. Further, we observed increase in the linewidth of $X^d$ peaks with increase in temperature (Supporting Information section - VIII), which is attributed to enhanced interaction of defect-bound excitons with phonons at higher temperatures. We also note that with increasing temperature, L peak quenches faster than $X^d$, indicating different physical origins and disassociation energies.

Second-order autocorrelation measurement is a pivotal tool for determining the single-photon nature and purity of an emitter. The second-order autocorrelation function ($g^{(2)}(\tau)$) for a two-level system at finite delay time ($\tau$) is defined as -

$$g^{(2)}(\tau) = 1 - ae^{-\tau/\tau_0}$$

Here, a and $\tau_0$ are fitting constant and lifetime of the emitter, respectively. The value of the $g^{(2)}(\tau)$ function at zero time delay ($\tau = 0$) is a critical metric for quantifying the purity of an emitter. For single photon emission, $g^{(2)}(0) < 0.5$, and for ideal SPEs, $g^{(2)}(0) = 0$. The second-order autocorrelation measurement of sharp peaks was performed using a 532 nm CW laser in a custom-built Hanbury Brown and Twiss (HBT) setup (see Methods). Figure 3d shows the graph of $g^{(2)}(\tau)$ plotted against the delay time ($\tau$). By fitting the data to the above equation, the extracted value of $g^{(2)}(0)$ is ~ 0.15 ± 0.05, which confirms the single-photon nature of the sharp defect peak.

Spectral jittering (average spectral shift of PL peaks over time) is a critical metric that measures the stability of SPEs. We recorded kinetic series of PL for over one hour, with each spectrum acquired for one second (Figure 3e). No blinking is observed in this sharp peak emitter. Then, the peak positions were obtained by fitting a Lorentzian function to each spectrum. The extracted peak positions were fitted to a Gaussian distribution, with jitter FWHM ~ 118 μeV. This jitter is nearly half of linewidth of the sharp peak (183 μeV), comparable to the highest-quality WSe$_2$ SPEs[9,13,14,31,32].

**Gated devices of nanoindented ML WSe$_2$ on SiO$_2$/Si**

A key advantage of nanoindentation on SiO$_2$/Si substrates is the ease of gated device fabrication, allowing measurement of the charge nature and doping dependence of SPEs. Firstly, the ML WSe$_2$ flakes were transferred near pre-patterned Ti/Au contacts on SiO$_2$/Si (patterned using electron beam lithography, see Methods). A graphite flake (~ 9 nm thickness) was transferred to bridge the ML WSe$_2$ with the metal contact pad. Contact resistance of TMDs with graphite is lower than with Ti/Au, which becomes important at low temperatures[33,34]. Then, AFM nanoindentation was performed as per the procedure described earlier. The schematic and corresponding AFM topography micrograph of the gated device featuring 20 indents (average depth = 203 nm ± 4 nm) are displayed in Figure 4a and 4b, respectively. The measured leakage current in the nanoindented sample is < 2 nA for 20 V gate voltage at room temperature, which is within the tolerance regime and is comparable to the leakage current of pristine samples (see Methods, and Supporting Information section - IX).

PL spectra were acquired from nanoindentation sites under varying gate voltages to investigate the gating response of ML WSe$_2$ SPEs (Figure 4c and Supporting Information section - X). At zero gate bias, prominent sharp peaks ($X^d$) were observed, superimposed on a broad background emission (L peak).

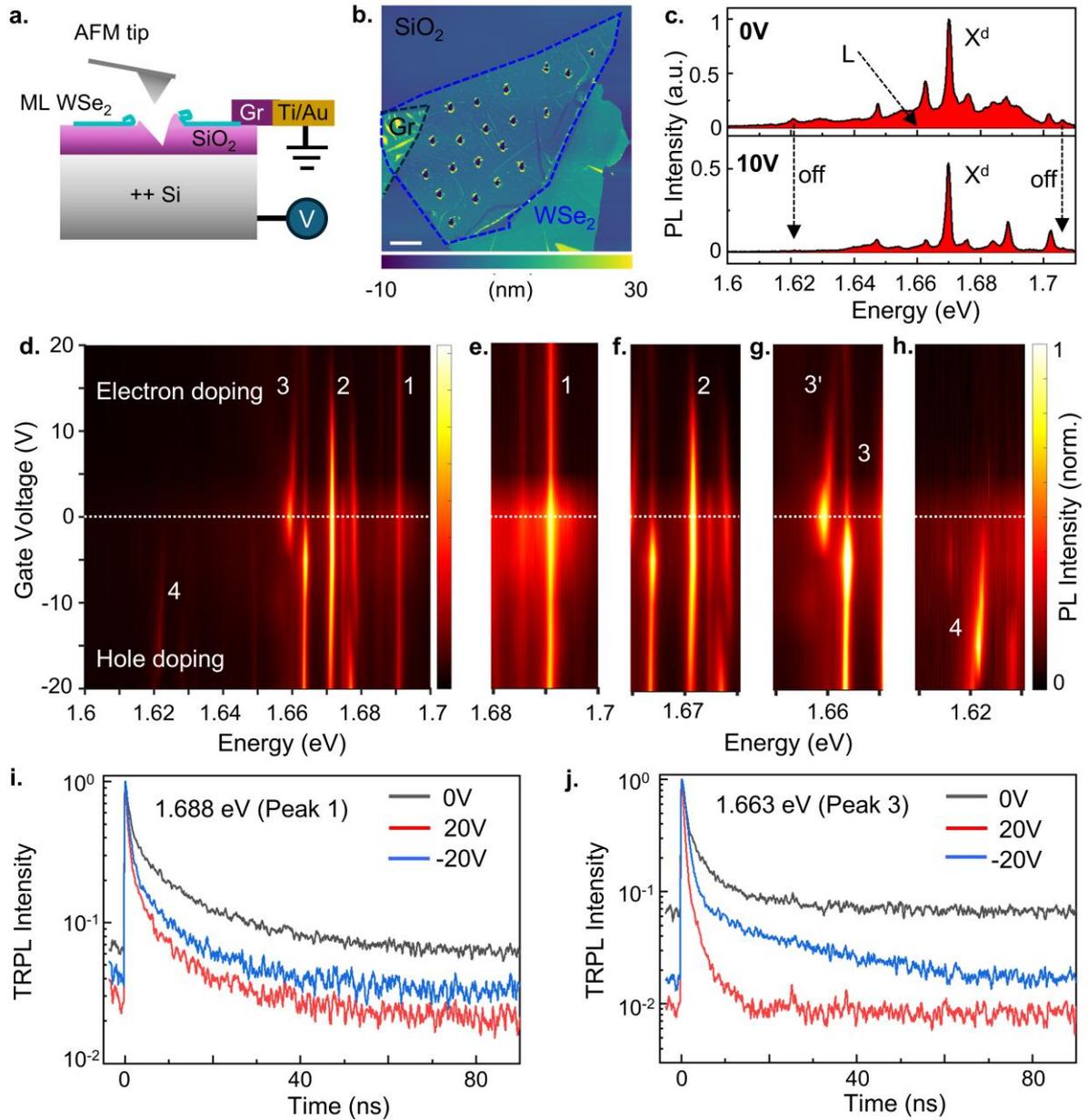

**Figure 4: Gate-voltage dependent PL and TRPL of nanoindented ML WSe$_2$ on SiO$_2$/Si substrate.** Schematics and AFM topography micrograph of a gated nanoindented ML WSe$_2$ device are shown in (a) and (b), respectively. Average indent depth is ∼ 203 nm. Horizontal scale bar in (b) is 3 μm. (c) Representative PL spectra acquired from nanoindented sites at 0 V and 10 V applied gate voltages. The PL spectra are normalized to maximum of 0 V PL data. (d) Surface plot showing evolution of sharp (X$^d$) peaks and L peak PL with applied gate voltage. (e-h) Surface plot of individual sharp peaks highlighting different charge nature. For (e-g), the PL spectra are normalized to maximum PL intensity of the respective peak at 0 V. For (g), the PL spectra is normalized to maximum PL intensity of the peak

at -15 V. **(i, j)** Gate-voltage dependent TRPL of sharp defect peaks ($X^d$) with energies 1.688 eV (Peak 1) and 1.663 eV (Peak 3).

Applying a positive bias of 10 V (electron doping), a few sharp peaks are turned off (highlighted in black dashed arrows in Figure 4c). This quenching effect may arise from instability of weakly-bound excitons under electron-doping conditions, or saturation of the defect states by electron doping. Further, we observed significant quenching in the L peak intensity with electron and hole doping (Figure 4c and Figure S10 (a), Supporting Information section - X). The broad L peak was previously attributed to defect-bound excitonic states[25,35]. The suppression of this classical background emission can be caused by the instability of defect-bound excitonic states at higher applied bias[16], which will enhance the purity of SPEs[10,16,36].

To further elucidate the stability and charge nature of the sharp emission peaks, gate-voltage dependent PL spectra were plotted as surface maps (Figure 4d - h). Figures 4e - h present normalized PL maps of four sharp peaks (1 - 4) under electron and hole doping conditions. The raw PL spectra used for creating surface maps is plotted in Figure S10 (b), Supplementary Information section - X. Peak 1 shows a slight initial quenching and then maintains a nearly constant PL intensity under both electron and hole doping, along with reduction of L peak background. Peak 2 remains largely unaffected under hole doping but is quenched under positive gate bias, indicating its instability under electron doping. Peak 3 emerges under hole doping, in contrast, peak 3' is completely suppressed under hole doping and gradually reduces under electron doping. Additional sharp emission peaks (peak 4) emerge exclusively under hole-doping conditions (Supporting Information section - X), indicating the activation of additional emitters. This behavior is attributed to electrostatic gating driving holes into the local potential wells formed by nanoindentation, activating multiple localized emitters, or holes creating different charge states of the same emitter[16]. The asymmetry in the response of the emitters to hole and electron doping suggests a correlation with their charge configuration.

To understand the charge nature and physical origin of emitters, along with the effect of L peak, we performed gate-voltage dependent TRPL (Figure 4i, j). At 0 V, the sharp peaks that are superimposed on L peak show biexponential TRPL decay, with a nanosecond-fast decay component and long-lived (5 - 15 ns) second component. The PL data at 0 V and the corresponding TRPL fits for three sharp peaks are presented in Supporting Information Section - XI. In contrast, sharp peaks located away from the L peak show single exponential TRPL decay (Figure 3b and Supporting Information Section - VII), indicating a different origin c.f. sharp peaks overlapping with L peak. This distinct nature of the sharp peaks is further supported by their varying g - factors[37].

Upon applying gate voltage, contribution of the initial (fast) decay for peak 1 and peak 3 increases, and the corresponding decay time shortens (Supporting Information Section - XI). However, the steady-state PL decreases, indicating that the initial fast decay is related to charge state and charge-stability of SPE. The long-lived component and TRPL background reduce at both positive and negative gate voltages, primarily due to reduced intensity of the L peak. The overall lifetime decreases at higher gate voltages (Figure 4i, 4j, and Supporting Information Section - XI), while steady-state PL reduces. We note that peak 3 has an anomalous response at -20 V, due to persisting background on the hole doping side (Figure S10 (a), Supporting Information Section - X). Thus, gate-dependent TRPL measurements help us understand the distinct charge nature of SPEs and reduction of background emission, guiding the way towards development of high purity SPEs.

**Conclusion:**

Through nanoindentation using a sharp diamond AFM tip, this work demonstrated the spatially-deterministic writing of ML WSe$_2$ SPEs directly on rigid and photonic structure compatible SiO$_2$/Si substrates. SPE creation was enabled by development of a displacement-controlled indentation process allowing indent depths > 150 nm. Cryogenic (4 K) PL mapping experiments revealed a high yield of SPE formation (~ 76%), demonstrating the reproducibility of the process. The SPEs were found to originate from the lower periphery of the indent, likely associated with localized out-of-plane deformations of the torn flake edge around the pronounced substrate bulge created during nanoindentation. SPEs showed defect-bound nature in laser power-dependent PL and TRPL experiments. The SPEs remained active up to 120 K, extending the operational temperature range toward practical liquid-nitrogen temperature applications. The SPE demonstrated excellent spectral stability (minimal jitter), no blinking, and high single-photon purity, essential for quantum technologies. Finally, we found that the SPEs can be turned on off and additional emitters can be activated via application of external gate bias. Importantly, the SPE dynamics can be tuned via electrical gating depending upon their emission energies, crucial for charge control and high purity SPE emission.

The process can be generalized to other 2D semiconductors, depending on the necessity of emission energies. Furthermore, the process can be applied to 2D magnetic materials for creation of chiral SPEs, and to incorporate ferroelectric tunability of SPEs on 2D TMDs on ferroelectric thin films (such as hafnium oxide) that exhibit a hardness comparable to that of SiO$_2$. Together, these results showcase a robust and versatile platform for developing scalable SPE sources with tuneable properties for future quantum applications. We believe our AFM nanoindentation technique for writing SPEs on rigid and photonic structure compatible substrates will pave the way for scalable fabrication of SPEs.

**Methods/Experimental section**

**Sample preparation:** WSe$_2$ flakes were exfoliated from bulk crystal (2D semiconductors) using Nitto blue tape and transferred onto SiO$_2$ (285nm) /Si substate using gel pack PDMS (polydimethylsiloxane) stamp. Monolayers of WSe$_2$ were identified and confirmed using optical microscopy[38]. For fabrication of gated samples, the gate electrodes and contact pads were patterned using 2 - layer PMMA e-beam lithography onto SiO$_2$ (285nm)/Si substrates. Subsequently, Ti/Au (10/100 nm) films were deposited using e-beam evaporator on the developed samples, followed by metal lift-off and solvent cleaning. ML WSe$_2$ was then transferred onto SiO$_2$/Si substate with pre-fabricated Ti/Au contact pad using dry transfer method. The ML WSe$_2$ was then bridged to Ti/Au contact pad via a graphite flake, following which AFM indentation was performed. The leakage current of the device was measured by applying an electric voltage to the Si gate and keeping the graphite layer (contact) grounded. At room temperature, the leakage current of the indented sample is within the tolerance regime (< 2 nA for 10 V gate voltage, see the Supporting Information section – IX).

**AFM Nanoindentation:**

Nanoindentation was performed using a Park NX20 AFM system. An Adama NM-RC AFM probe with a sharp diamond tip, with manufacturer specified tip radius R = 10 ± 5 nm, nominal spring constant k = 350 N/m, and nominal resonance frequency $f_0$ = 850 kHz, was used. Displacement-controlled AFM nanoindentation was performed with Z-piezo displacement defined as per the target indent depth. The required Z-piezo displacement was calibrated by performing indentations for the target depth on the SiO$_2$/Si substrate. The loading/unloading rate and dwell time for each indentation cycle was kept constant at 0.5 μm/s and 0.1 s respectively. For each device, as a part of calibration, initially, indentations with varying Z-piezo displacements typically between 1 μm to 2 μm were performed on the regions of SiO$_2$/Si substrate not covered with the flake. Then imaging of the indents was performed with same probe in non-contact mode to obtain a correlation between Z-piezo displacement and corresponding indent depth. Finally, a Z-piezo displacement value corresponding to the desired indent depth was determined from this correlation to perform indentation on WSe$_2$ on SiO$_2$/Si substrate. Spacing between two adjacent indents on the WSe$_2$ flake was ~ 5 μm to ensure that the laser (spot size 1.3 μm) probes only one indented site during the cryogenic (4 K) PL measurements.

Topography of the indents created on the WSe$_2$ flake was measured in Bruker Dimension Icon AFM system with tapping mode using a NuNano Scout 150 HAR, high-aspect ratio sharp silicon probe with manufacturer specified values of tip radius $R = 5$ nm, nominal spring constant $k = 18$ N/m, and nominal resonance frequency $f_0 = 150$ kHz. A cantilever free amplitude of 300 mV with a setpoint of 150 mV to 170 mV were found to be optimal for simultaneous topography and phase-imaging of the indented site.

**Optical spectroscopy:**

For cryogenic (4 K) PL measurements, a 635 nm CW laser (Thor Labs) was focused on the sample cooled in Montana S50 Cryostation using a 50x objective (Mitutoyo, 0.45 NA). For acquiring PL spectrum, the emission from the sample was directed to the assembly of Andor CCD and spectrometer with 300 and 1200 lines/mm gratings. The spectral resolution for 1200 lines/mm grating is ~ 70 μeV. The laser was blocked before the spectrometer using high-performance long-pass filters. For TRPL measurement, a pulsed ultrafast laser (690 nm and 710 nm) excitation source of 100 fs pulse-width and 8 MHz repetition rate was used. PL signal decay over time was recorded using an assembly of single photon avalanche photodiodes (SPADs from MPD) and timing electronics (PicoHarp 300).

**Cryogenic (4 K) PL mapping:**

Cryogenic PL maps of Figure 2 were acquired in a Attocube AttoDry1000 closed cycle cryostat using piezoelectric scanners to move the sample. A 633 nm continuous-wave He-Ne laser, coupled to a single-mode fiber, serve as the excitation source. The excitation spot is focused on the sample down to the diffraction limit using a high numerical aperture (NA=0.82) microscope objective. The PL signal is coupled to another single mode fiber ensuring a confocal geometry. The detection fiber is then connected to a SPAD (Excelitas). Tuneable long-pass and short-pass filters are used to select the detection energy range. PL spectra are also acquired by connecting the detection fiber to a monochromator coupled with a charge-coupled device (CCD) camera.

**Second-order autocorrelation measurement:**

Second-order autocorrelation measurements were carried out using a custom optical setup comprising a closed-cycle, cryogen-free cryostat (AttoDry800, attocube systems AG), an XYZ nano positioner, and an LT APO objective (NA = 0.82). A 532 nm CW laser was used for excitation. Photon statistics were recorded with a fiber-based HBT interferometer, where coincidence events from two Si APDs (Excelitas SPCM-AQRH16) were detected on synchronized channels with an overall system timing jitter of ~ 600 ps. The SPE peak was isolated from the total emission using a tuneable bandpass filter.

ASSOCIATED CONTENT

**Supporting Information:**

AFM force-controlled nanoindentation of monolayer $WSe_2$, Z-piezo displacement calibration on a $SiO_2$/Si substrate, depth-dependent nanoindentation of monolayer $WSe_2$, reference PL spectra for cryogenic (4 K) PL maps in different energy ranges, indentation of monolayer $WSe_2$ using nanoindenter tool, analysis of laser power-dependent PL data, analysis of TRPL data of the $X^d$ and L peak, evolution of linewidth and energy position of $X^d$ peak with temperature, leakage current measurement of gated device after indentation, gate-voltage dependent PL of nanoindented ML $WSe_2$, gate-voltage dependent TRPL of nanoindented ML $WSe_2$.

The following files are available free of charge: Supplementary_information.pdf

AUTHOR INFORMATION:

**Corresponding Author:**

*Akshay Naik (anaik@iisc.ac.in), Akshay Singh (aksy@iisc.ac.in)

**Author Contributions:**

AKD, SJ, AN, and AS developed the experimental framework. AKD prepared exfoliated samples. SJ conceived, optimized, and conducted the AFM nanoindentation experiments on the exfoliated samples and performed all associated AFM data analysis, with supervision from AN. AKD performed optical characterization of indented samples with assistance from MPS. MPS fabricated electrical contact pads, with guidance from AN. AKD and Hardeep performed gate voltage dependent PL experiments. TG, CCS, and SR performed cryogenic PL mapping and jittering experiments with guidance from CR, XM, and AS. IDP, YW, AKD, SS, and MPS performed second-order autocorrelation measurements with guidance from SK and AS. AKD performed the PL and TRPL data analysis with guidance from AS. AKD, SJ, AN, and AS discussed and drafted the manuscript, with contributions from all authors. All authors have given approval to the final version of the manuscript. †AKD and †SJ contributed equally.

**Notes:**

The authors declare that a patent application related to this work is pending.

**Data Availability :**

All data is available upon reasonable request.

## ACKNOWLEDGMENTS

AS acknowledges funding from Department of Science and Technology (DST) Nanomission CONCEPT grant (NM/TUE/QM-10/2019), DST National Quantum Mission (DST/QTC/NQM/QC/2024/1), and MoE-STARS (STARS-2/2023-0265). AS and CR acknowledge funding from CEFIPRA grant (2025/7304-4). AS and AN acknowledge financial support from the Quantum Research Park, which is funded by the Government of Karnataka, India. AKD and SJ acknowledge Prime Minister's Research Fellowship (PMRF). The authors acknowledge Prof. Praveen Kumar and Anjali Gawande, Department of Materials Engineering, Indian Institute of Science Bangalore for help in the instrumented nanoindentation measurements. The authors acknowledge Prof. Chandni Usha, Rohit Singh Bhandari, and Suvronil Datta, Department of Instrumentation and Applied Physics, Indian Institute of Science, for help in wire bonding of the gated devices. The authors also acknowledge the National Nanofabrication Centre (NNfC) and Micro Nano Characterization Facility (MNCF) at Centre for Nano Science and Engineering (CeNSE), Indian Institute of Science Bangalore. SK acknowledges funding from DST Nano Mission grant (DST/NM/TUE/QM-2/2019) and the matching grant from IIT Goa. IDP thanks the Council of Scientific & Industrial Research (CSIR), New Delhi, for the doctoral fellowship.

## ABBREVIATIONS

SPEs, single photon emitters; 2D, two-dimensional; ML, monolayer, AFM, atomic force microscopy; PL, photoluminescence; TRPL, time-resolved PL; TMDs, transitional metal dichalcogenides; PMMA, polymethyl methacrylate; PDMS, polydimethylsiloxane; ZPL, zero phonon line; PSB, phonon side band; FWHM, full width at half maximum; CCD, charge coupled device.

**Supporting Information**

**Gate-tuneable single-photon emitters in WSe$_2$ monolayer created via AFM nanoindentation on rigid SiO$_2$/Si substrates**

*[1†]Ajit Kumar Dash, [2†]Sanket Jugade, [2]Manavendra Pratap Singh, [1]Hardeep, [3]Tilly Guyot, [3]Cora Crunteanu-Stanescu, [4]Indrajeet Dhananjay Prasad, [4]Yunus Waheed, [4]Sumitra Shit, [3]Sébastien Roux, [4]Santosh Kumar, [3]Cedric Robert, [3,5]Xavier Marie, [2*]Akshay Naik, [1*]Akshay Singh*

[1]Department of Physics, Indian Institute of Science, Bengaluru, Karnataka 560012, India

[2]Centre for Nanoscience and Engineering, Indian Institute of Science, Bengaluru, Karnataka 560012, India

[3]Université de Toulouse, INSA-CNRS-UPS, LPCNO, 135 Avenue de Rangueil, 31077 Toulouse, France

[4]School of Physical Sciences, Indian Institute of Technology Goa, Ponda, 403401, Goa, India

[5]Institut Universitaire de France, 75231 Paris, France

† Authors contributed equally to the work

*Corresponding authors: anaik@iisc.ac.in, aksy@iisc.ac.in

(I) AFM force-controlled nanoindentation of monolayer (ML) WSe$_2$ on SiO$_2$/Si substrate

(II) Z-piezo displacement calibration on a SiO$_2$/Si substrate

(III) Depth-dependent nanoindentation of ML WSe$_2$ on SiO$_2$/Si substrate

(IV) Reference PL spectra for cryogenic (4 K) PL maps in different energy ranges

(V) Indentation of ML WSe$_2$ on SiO$_2$/Si substrate using instrumented nanoindenter

(VI) Analysis of laser power-dependent PL data

(VII) Analysis of TRPL data of the $X^d$ and L peak

(VIII) Evolution of linewidth and energy position of $X^d$ peak with temperature

(IX) Leakage current measurement of gated device after indentation

(X) Gate-voltage dependent PL of nanoindented ML WSe$_2$.

(XI) Gate-voltage dependent TRPL of nanoindented ML WSe$_2$.

**(I) AFM force-controlled nanoindentation of monolayer (ML) WSe$_2$ on SiO$_2$/Si substrate**

Force-controlled AFM nanoindentation was performed by applying a specified setpoint force (kept below the 10 V saturation of photodiode) with a Z-piezo ramping speed of 0.2 µm/s and dwell time of 0.1 s. The corresponding force value, expressed in nanonewtons or micronewtons, was calculated through a post-indentation calibration of deflection sensitivity and spring constant to preserve tip quality. The deflection sensitivity was obtained by measuring the slope of the retrace segment of the force-distance curve acquired on a hard reference sample (hBN flakes on SiO$_2$/Si). The spring constant was calibrated using the manufacturer's Laser doppler Vibrometer-based calibration formula, $k = AQf^{1.3}$, where, $A$ is a manufacturer-provided constant different for each tip box, $f$ and $Q$ are the resonance frequency and quality factor of the AFM probe obtained from the amplitude vs. frequency tuning curve.

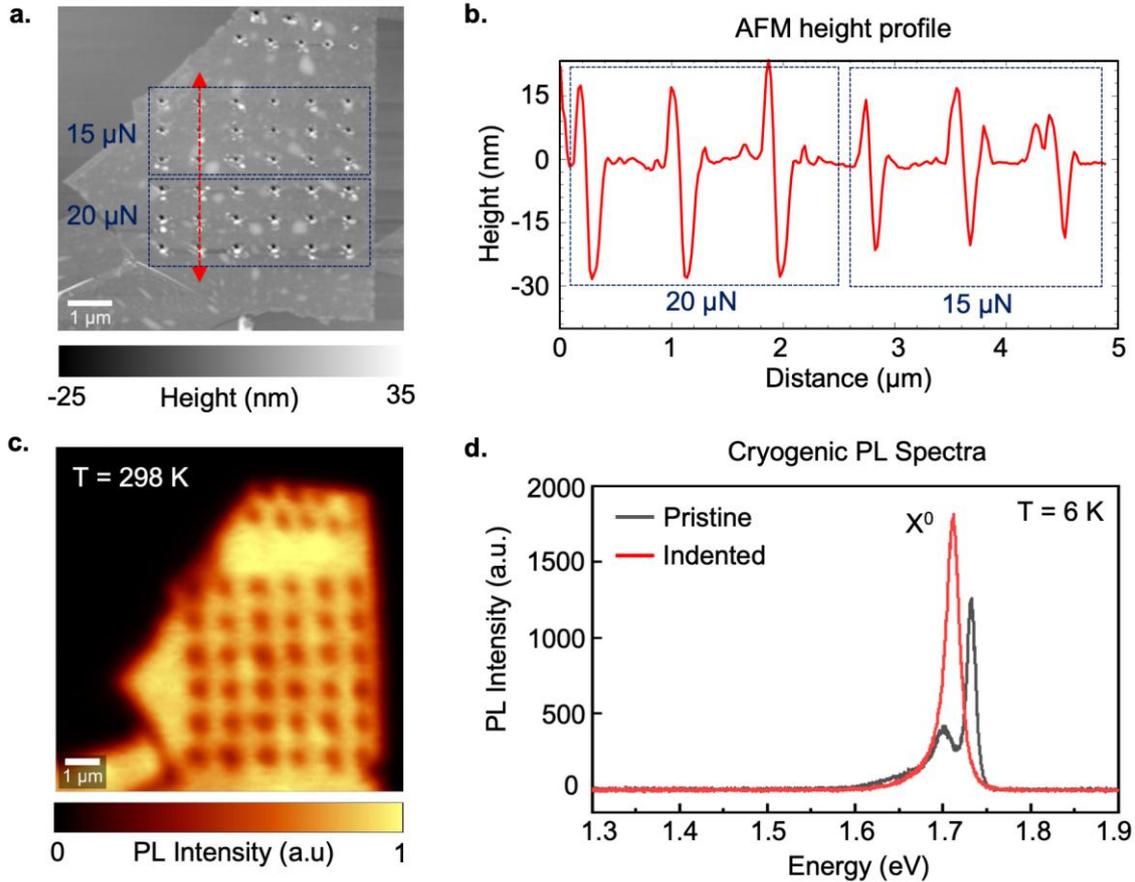

**Figure S1**. **Force-controlled nanoindentation and photoluminescence (PL) modulation in monolayer (ML) WSe$_2$ on SiO$_2$/Si substrate.** **(a)** AFM topography micrograph of a ML WSe$_2$ flake on SiO$_2$/Si substrate after AFM nanoindentation. An array of 6 × 6 indents was created using a force-controlled approach, with the bottom three rows indented at 20 μN force and the top three rows at 15 μN. **(b)** Topography line profile along the dashed red line in (a), showing the 2D profile and depth of the indents, with the maximum depth reaching ∼ 30 nm for the highest applied force (20 μN), corresponding to the upper limit of the measurable force in the AFM system and ∼ 20 nm for the applied force of 15 μN. **(c)** Room-temperature PL map of the flake, showing pronounced PL quenching at the indented locations due to AFM tip shear-force induced sculpting of WSe$_2$. This behaviour is consistent with observations from a previous nanoindentation study on WSe$_2$ on SiO$_2$/Si substrate using an instrumented indenter[1]. **(d)** Cryogenic (6 K) PL spectra acquired from a pristine flake and nanoindented flake, showing PL modulation and no emergence of sharp defect peaks due to force-controlled nanoindentation.

**(II) Z-piezo displacement calibration on a SiO₂/Si substrate**

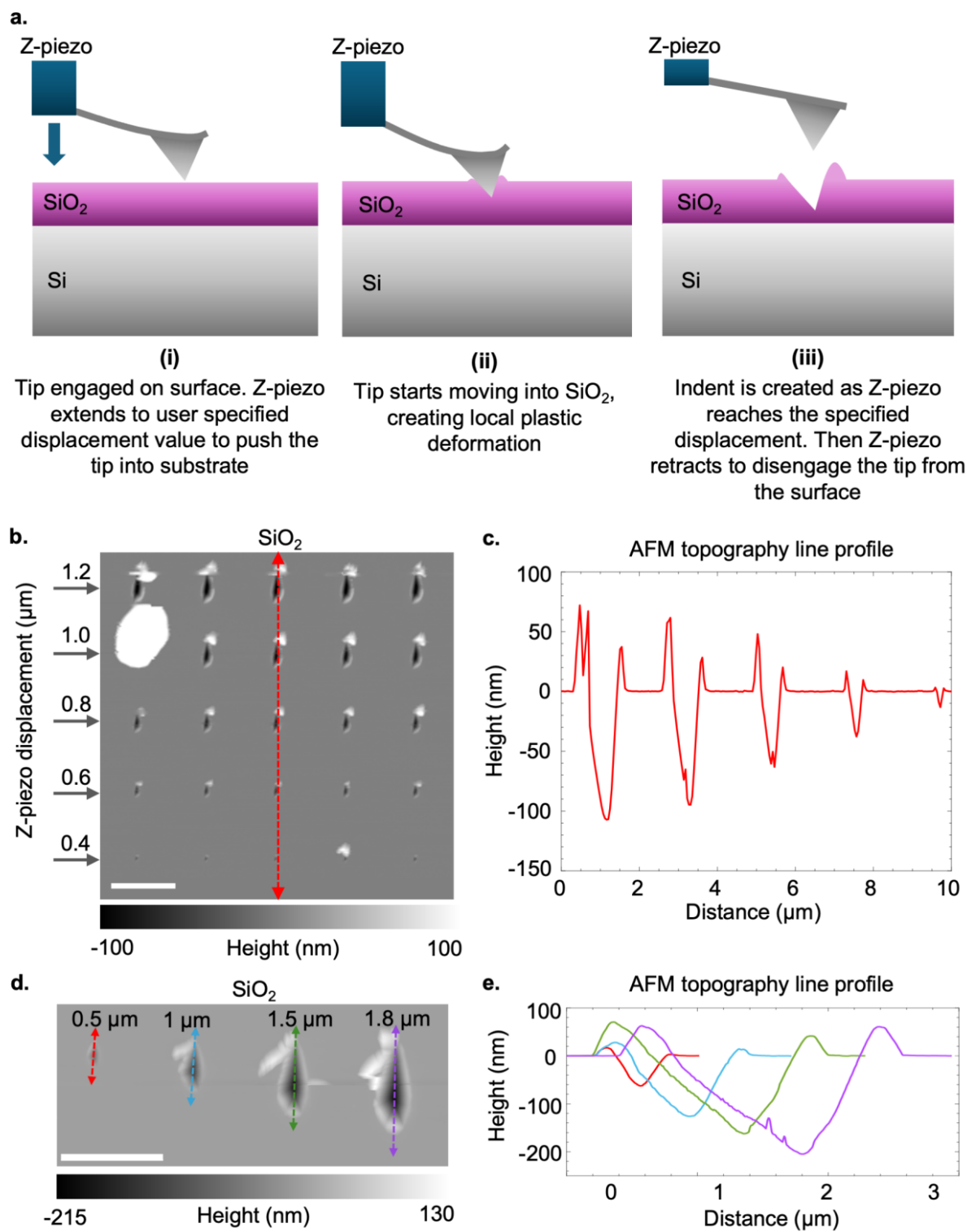

**Figure S2. Z-piezo displacement calibration on a SiO₂/Si substrate. (a)** Schematic of the displacement-controlled AFM nanoindentation on SiO₂/Si substrate. **(b)** AFM topography of an array of 5 × 5 indents created by systematically increasing the Z-piezo displacement value. Each row contains

five indents with consistent shape and depth made with constant Z-piezo displacements of 0.4 μm, 0.6 μm, 0.8 μm, 1.0 μm, and 1.2 μm, respectively. Horizontal scale bar is 2 μm. **(c)** Topography line profile along the marked red line in (b), showing the 2D cross-sectional profiles and corresponding depths of the indents. **(d)** AFM topography of indents made on a separate $SiO_2$/Si substrate using increased Z-piezo displacements ranging from 0.5 μm to 1.8 μm, illustrating the need for higher displacements to achieve indent depths greater than ~ 150 nm for SPE generation. Horizontal scale bar is 2 μm. **(e)** Line profiles along the individual indents in (d), demonstrating the progressive increase in depth with higher Z-piezo displacement values. These measurements also highlight the necessity of performing Z-piezo displacement calibration on each new sample, as the same Z-piezo displacement can result in different indent depths depending on variations in sample mechanical properties, flatness, cantilever stiffness, and tip radius and geometry. Accurate calibration ensures reproducibility in indent shape and depth across multiple devices.

## (III) Depth-dependent nanoindentation ML WSe$_2$ on SiO$_2$/Si substrate

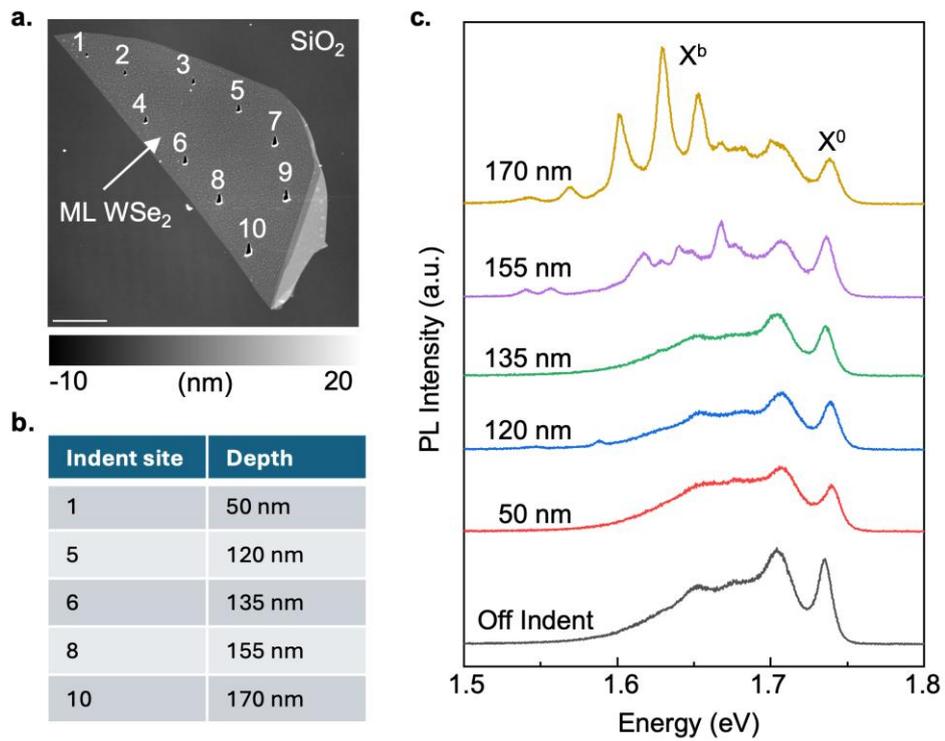

**Figure S3. Depth-dependent nanoindentation and cryogenic (6 K) PL of ML WSe$_2$ on SiO$_2$/Si substrate. (a)** AFM topography micrograph of a ML WSe$_2$ flake on a SiO$_2$/Si substrate featuring ten indents with increasing depths, achieved by progressively increasing the Z-piezo displacement. Horizontal scale bar is 5 μm. **(b)** The corresponding depths of nanoindented sites 1, 5, 6, 8, and 10 as extracted from the AFM topography micrographs. **(c)** Cryogenic (6 K) PL spectra acquired from off-indent site and indented sites 1, 5, 6, 8, and 10 revealing the correlation between indentation depth and sharp defect peak formation.

**(IV) Reference PL spectra for cryogenic (4 K) PL maps in different energy ranges**

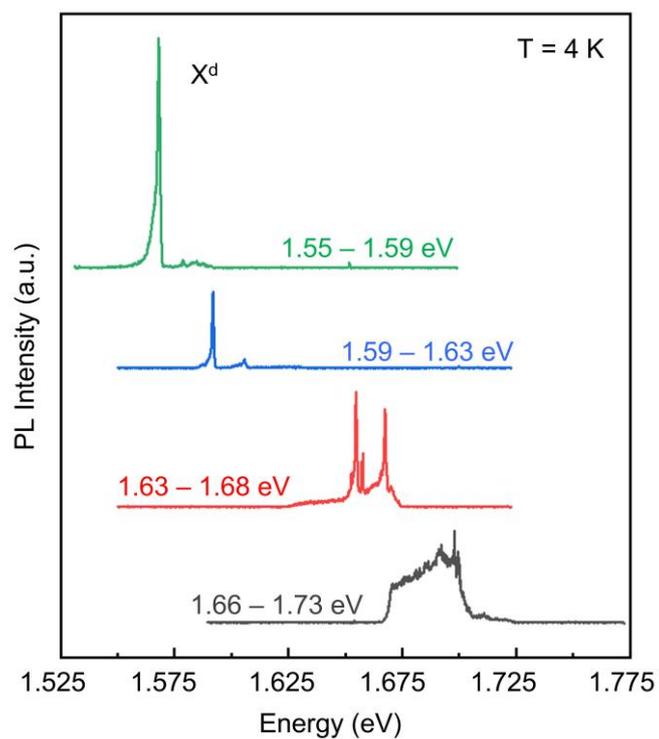

**Figure S4.** Reference PL spectra for cryogenic (4 K) PL maps in different energy ranges.

**(V) Indentation of ML WSe$_2$ on SiO$_2$/Si substrate using instrumented nanoindenter**

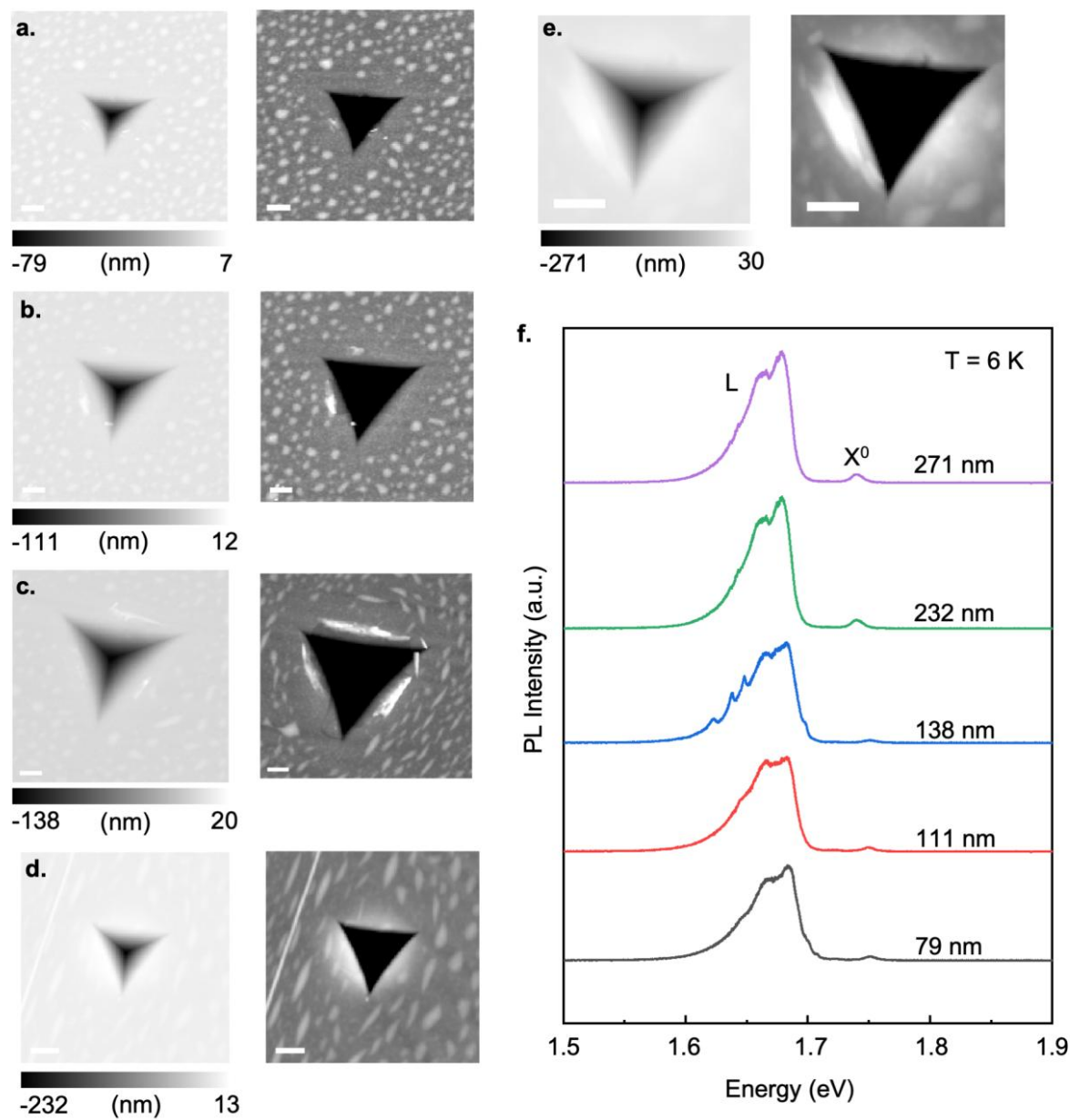

**Figure S5. Indentation of ML WSe$_2$ on SiO$_2$/Si substrate with an instrumented nanoindenter. (a-e)** AFM topography micrographs of indents with varying depths created by displacement-controlled indentation on ML WSe$_2$ flake using an instrumented nanoindenter equipped with a Berkovich tip (nominal end radius ~ 100 nm). The indentation depths are 79 nm, 111 nm, 138 nm, 232 nm, and 271 nm. To enhance visual contrast, corresponding AFM images for each of the micrographs in (a-e) are shown with a reduced vertical scale, highlighting the flake tearing within the indent more distinctly. All the horizontal scale bars in (a-e) are 200 nm. The indents show a symmetric profile with a uniformly

distributed SiO$_2$ pile-up (bulge) around their perimeters. For the deepest indent, with a depth of 271 nm, the bulge height is ~30 nm. This substrate bulging produced by the instrumented indenter is significantly lower than that observed in AFM-nanoindentation created indents, where a minimum required indent depth of 150 nm for sharp PL emission results in an SiO$_2$ bulge height of about 81 nm. **(f)** Cryogenic (6 K) PL spectra acquired from each of the indented sites. No sharp peaks (X$^d$) were observed.

## (VI) Analysis of laser power-dependent PL data

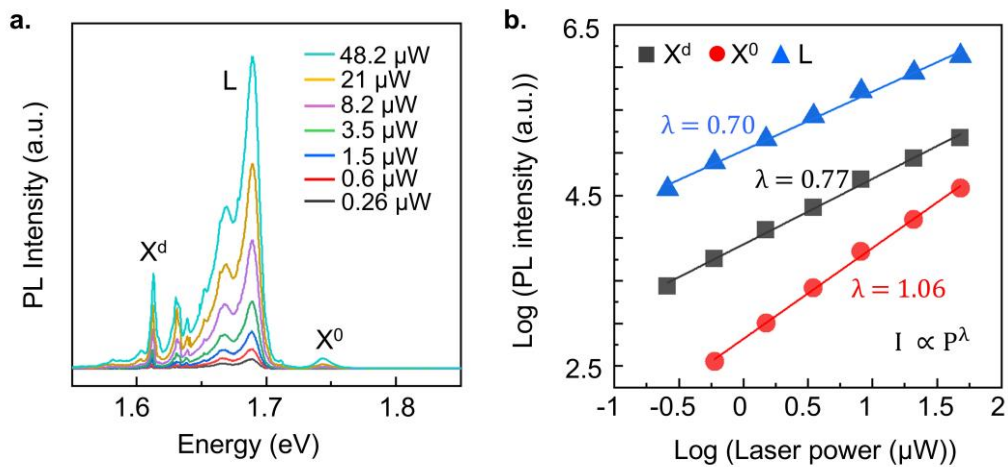

**Figure S6. Laser power-dependent PL of sharp (X$^d$), neutral (X$^0$) excitons, and broad classical background (L) peak. (a)** Raw cryogenic (4 K) PL spectra of nanoindented ML WSe$_2$ acquired at different laser powers. **(b)** Power coefficient ($\lambda$) analysis of X$^d$, X$^0$, and L peaks. The X$^d$ and L peaks show sublinear and X$^0$ show linear dependence on laser power.

## (VII) Analysis of TRPL data of the $X^d$ and L peak

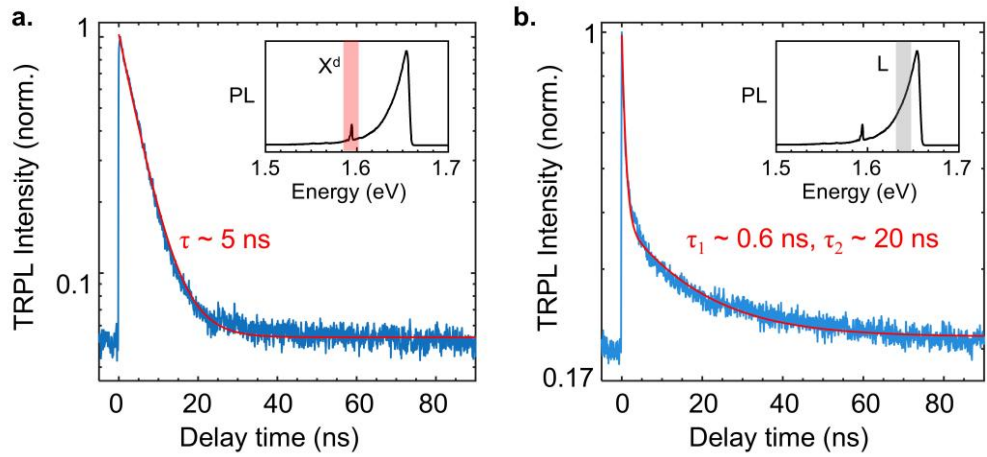

**Figure S7. Time resolved photoluminescence (TRPL) spectra of sharp peak ($X^d$) and L peak.** Single exponential and biexponential decay function fitting of $X^d$ and L peak are shown in **(a)** and **(b)**, respectively. The representative PL spectrum acquired from the nanoindented site is shown in the insets of panels (a) and (b). The shaded region in the insets indicate the spectral bands selected for TRPL measurements.

| Peak | $a_0$ | $a_1$ | $\tau_1$ (ns) | $a_2$ | $\tau_2$ (ns) |
|---|---|---|---|---|---|
| $X^d$ | 0.05 | $0.95 \pm 0.01$ | $5.1 \pm 0.1$ | - | - |
| L | 0.2 | $0.62 \pm 0.01$ | $0.9 \pm 0.1$ | $0.16 \pm 0.01$ | $20.1 \pm 0.1$ |

**Table S1:** Extracted parameters for single and biexponential fit of the sharp peak ($X^d$) and L peak, respectively.

## (VIII) Evolution of linewidth and energy position of $X^d$ peak with temperature

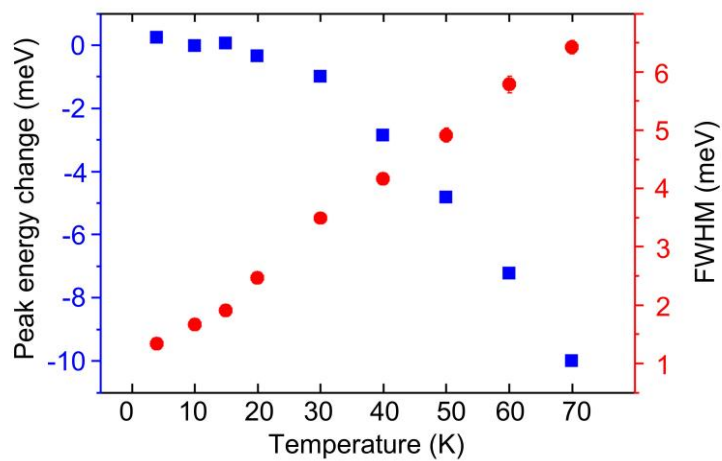

**Figure S8. Evolution of peak energy and full width at half maximum (FWHM) of sharp peak ($X^d$) with temperature.** The change in peak energy (with respect to the peak energy at 4 K) and variation of FWHM of sharp defect peak ($X^d$) with temperature are plotted in left and right axis, respectively.

## (IX) Leakage current measurement of gated device after indentation

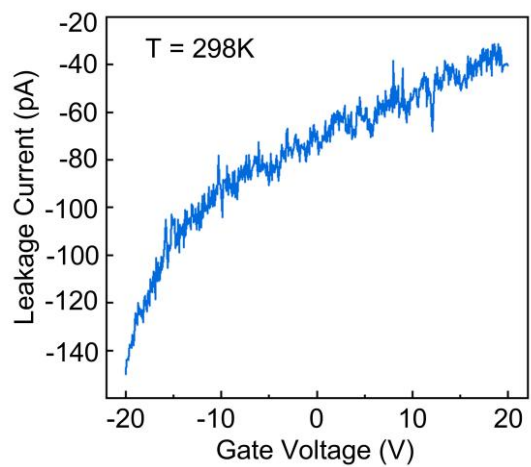

**Figure S9.** Leakage current of the gated sample after nanoindentation, measured at room temperature.

**(X) Gate-voltage dependent PL of nanoindented ML WSe$_2$.**

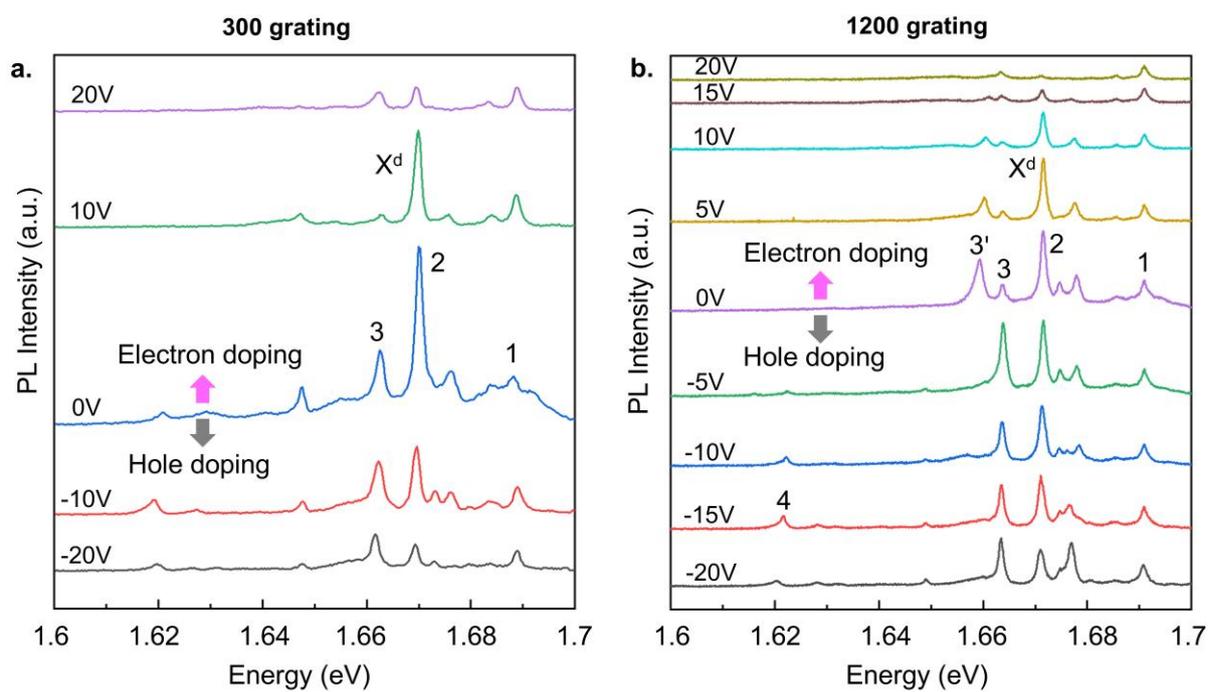

**Figure S10. (a, b)** Gate-voltage dependent PL of sharp defect peaks (X$^d$) acquired using 300 lines/mm and 1200 lines/mm grating, respectively.

**(XI) Gate-voltage dependent TRPL of nanoindented ML WSe$_2$.**

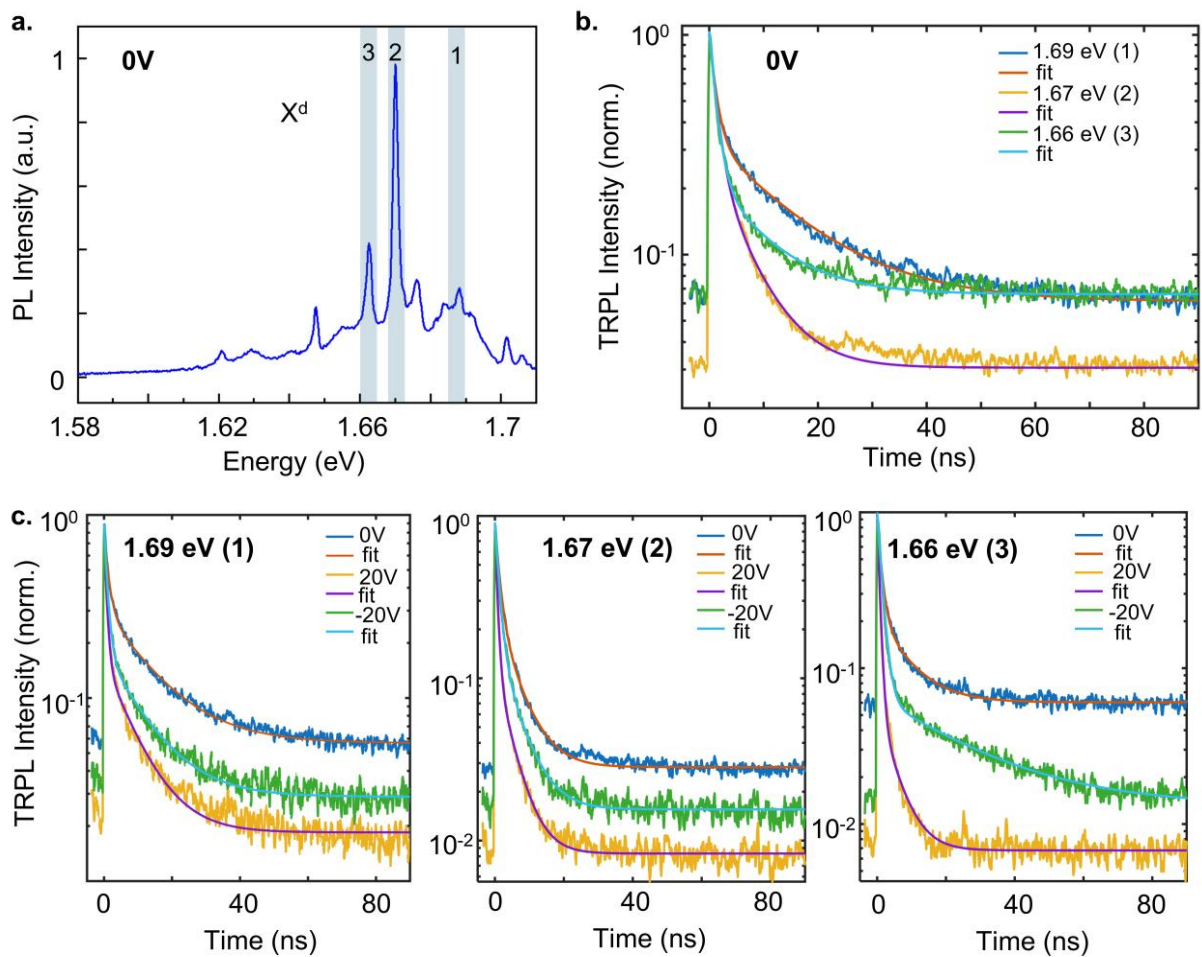

**Figure S11.** Cryogenic (4 K) PL and TRPL of three sharp peaks (1, 2, 3) at 0 V are shown in **(a)** and **(b)**, respectively. **(c)** Gate-voltage dependent TRPL of sharp peaks 1 - 3.

| Gate-voltage | $a_0$ | $a_1$ | $\tau_1$ (ns) | $a_2$ | $\tau_2$ (ns) |
|---|---|---|---|---|---|
| 20V | 0.02 | 0.81 ± 0.01 | 0.7 ± 0.1 | 0.17 ± 0.01 | 9.2 ± 0.1 |
| 0V | 0.06 | 0.67 ± 0.01 | 1.2 ± 0.1 | 0.28 ± 0.01 | 14.2 ± 0.1 |
| -20V | 0.03 | 0.84 ± 0.01 | 1.0 ± 0.1 | 0.17 ± 0.01 | 12.1 ± 0.2 |

**Table S2**: Extracted parameters for biexponential fit of the sharp peak 1 (1.69 eV).

| Gate-voltage | $a_0$ | $a_1$ | $\tau_1$ (ns) | $a_2$ | $\tau_2$ (ns) |
|---|---|---|---|---|---|
| 20V | 0.01 | 0.90 ± 0.01 | 0.7 ± 0.1 | 0.14 ± 0.01 | 4.4 ± 0.1 |
| 0V | 0.03 | 0.74 ± 0.01 | 1.3 ± 0.1 | 0.24 ± 0.01 | 6.1 ± 0.1 |
| -20V | 0.02 | 0.85 ± 0.01 | 1.1 ± 0.1 | 0.18 ± 0.01 | 5.5 ± 0.1 |

**Table S3:** Extracted parameters for biexponential fit of the sharp peak 2 (1.67 eV).

| Gate-voltage | $a_0$ | $a_1$ | $\tau_1$ (ns) | $a_2$ | $\tau_2$ (ns) |
|---|---|---|---|---|---|
| 20V | 0.01 | 1.02 ± 0.01 | 0.7 ± 0.1 | 0.06 ± 0.01 | 4.7 ± 0.3 |
| 0V | 0.07 | 0.81 ± 0.01 | 1.2 ± 0.1 | 0.17 ± 0.01 | 9.2 ± 0.2 |
| -20V | 0.02 | 1.02 ± 0.01 | 1.3 ± 0.1 | 0.06 ± 0.01 | 23.1 ± 0.7 |

**Table S4:** Extracted parameters for biexponential fit of the sharp peak 3 (1.66 eV).